\documentclass[conference]{IEEEtran}
\IEEEoverridecommandlockouts

\usepackage[a4paper, total={184mm,239mm}]{geometry}

\usepackage[numbers,square,compress]{natbib}

\usepackage{fancyhdr}

\usepackage{tikzsymbols}
\usepackage{array}
\usepackage{tablefootnote}
\usepackage{threeparttable}
\usepackage{upgreek}
\usetikzlibrary{shapes}
\usepackage{glossaries} 

\usepackage{xcolor}
\usepackage{ifthen}

\usepackage{makecell}
\usepackage{booktabs}

\usepackage{siunitx}

\usepackage{hhline}
\usepackage{float}
\usepackage{algorithm}
\usepackage{datetime}
\usepackage{soul}
\usepackage{support-caption}

\usepackage{amsmath}
\usepackage{amssymb}
\usepackage{cases}

\usepackage{marginnote} 
\let\marginpar\marginnote
\usepackage{hyperref}

\usepackage{algorithmic}
\usepackage{graphicx}
\usepackage{textcomp}
\usepackage{todonotes} 

\usepackage{textcomp}
\usepackage[T1]{fontenc}

\usepackage{balance}
\usepackage{flushend}

\usepackage{tcolorbox}

\usepackage{mathtools}

\usepackage[framemethod=tikz]{mdframed}
\usepackage{setspace}
\usepackage{blindtext}

\usepackage{imakeidx}

\usepackage{multicol}
\usepackage{multirow}
\title{Multicols Demo}
\usepackage{tabularx}

\usepackage{enumitem}

\let\oldmarginnote\marginnote

\renewcommand{\marginnote}[2][rectangle,draw,fill=blue!40,rounded corners]{%
        \oldmarginnote{%
        \tikz \node at (0,0) [#1]{#2};}%
        }

\definecolor{lightyellow}{rgb}{0.980, 0.956, 0.623}
\sethlcolor{lightblue}

\definecolor{amber}{rgb}{1.0, 0.49, 0.0}
\definecolor{awesome}{rgb}{1.0, 0.13, 0.32}
\definecolor{dollarbill}{rgb}{0.52,0.73,0.4}
\definecolor{moegi}{rgb}{0.357, 0.537, 0.188}
\definecolor{burgundy}{rgb}{0.5, 0.0, 0.13}
\definecolor{ballblue}{rgb}{0.13, 0.67, 0.8}
\definecolor{ups-truck}{rgb}{0.53, 0.28, 0.21}
\definecolor{airforceblue}{rgb}{0.36, 0.54, 0.66}
\definecolor{cadmiumgreen}{rgb}{0.0, 0.42, 0.24}
\definecolor{darkcyan}{rgb}{0.0, 0.55, 0.55}
\definecolor{caribbeangreen}{rgb}{0.0, 0.8, 0.6}
\definecolor{flamingopink}{rgb}{0.99, 0.56, 0.67}
\definecolor{jazzberryjam}{rgb}{0.65, 0.04, 0.37}
\definecolor{mediumpersianblue}{rgb}{0.0, 0.4, 0.65}
\definecolor{coolblack}{rgb}{0.0, 0.18, 0.39}
\definecolor{bleudefrance}{rgb}{0.19, 0.55, 0.91}
\definecolor{ao}{rgb}{0.0, 0.0, 1.0}
\definecolor{babyblueeyes}{rgb}{0.63, 0.79, 0.95}
\definecolor{darkwarmgray}{rgb}{0.2,0,0}
\definecolor{lightblue}{rgb}{0.980, 0.956, 0.623}

\newcommand*\circled[1]{\tikz[baseline=(char.base)]{
            \node[shape=circle,draw,inner sep=0pt,fill=black, text=white] (char) {#1};}}

\newcommand\fig[1]{Fig.~{#1}\xspace}

\newcommand\sect[1]{Section~{#1}\xspace}

\newcommand\eque[1]{Equation~{#1}\xspace}

\newcommand{\cim}{CIM\xspace}
\newcommand{\cimlong}{Computation-In-Memory\xspace}

\newcommand{\mac}{MAC\xspace}
\newcommand{\maclong}{Multiply-and-Accumulate\xspace}
\newcommand{\vmm}{VMM\xspace}
\newcommand{\vmmlong}{Vector-Matrix-Multiplication\xspace}
\newcommand{\mmm}{MMM\xspace}
\newcommand{\mmmlong}{Matrix-Matrix-Multiplication\xspace}

\newcommand{\puma}{PUMA\xspace}

\newcommand{\sotalong}{State-of-the-Art\xspace}
\newcommand{\sota}{SotA\xspace}

\newcommand{\gpu}{GPU\xspace}

\newcommand{\nnlong}{Neural Network\xspace}
\newcommand{\nn}{NN\xspace}
\newcommand{\dnnlong}{Deep Neural Network\xspace}
\newcommand{\dnn}{DNN\xspace}

\newcommand{\cnn}{CNN\xspace}
\newcommand{\bnnlong}{Binary Neural Network\xspace}
\newcommand{\bnn}{BNN\xspace}

\newcommand{\daclong}{digital to analog converter\xspace}
\newcommand{\dac}{DAC\xspace}

\newcommand{\adclong}{analog to digital converter\xspace}
\newcommand{\adc}{ADC\xspace}

\newcommand{\sa}{SA\xspace}
\newcommand{\minusone}{\{-1, 1\}\xspace}
\newcommand{\zeroone}{\{0, 1\}\xspace}

\newcommand{\pcm}{PCM\xspace}
\newcommand{\pcmmatlongcapital}{Phase Change Material\xspace}

\newcommand{\epcmlongsmall}{electronic phase change memory\xspace}
\newcommand{\epcm}{ePCM\xspace}

\newcommand{\opcmlongsmall}{optical phase change memory\xspace}
\newcommand{\opcm}{oPCM\xspace}

\newcommand{\cmos}{CMOS\xspace}
\newcommand{\reramlongsmall}{resistive random-access memory\xspace}
\newcommand{\reram}{ReRAM\xspace}

\newcommand{\gatexnorop}{XNOR\xspace}

\newcommand{\popcount}{Popcount\xspace}
\newcommand{\gatexnoroppluspopcount}{XNOR+Popcount\xspace}

\newcommand{\wdmlong}{wavelength division multiplexing\xspace}
\newcommand{\wdm}{WDM\xspace}
\newcommand{\voalong}{variable optical attenuator\xspace}
\newcommand{\voa}{VOA\xspace}

\newcommand{\tialong}{transimpedance amplifiers\xspace}
\newcommand{\tia}{TIA\xspace}
\newcommand{\ecore}{ECore\xspace}
\newcommand{\vcore}{VCore\xspace}

\newcommand{\baseline}{Baseline-ePCM\xspace}
\newcommand{\gpubaseline}{Baseline-GPU\xspace}

\newcommand{\bnnopcmacc}{EinsteinBarrier\xspace}
\newcommand{\effmap}{TacitMap\xspace}
\newcommand{\epcmeffmap}{TacitMap-ePCM\xspace}

\newcommand{\pcsabinarymapping}{CustBinaryMap\xspace}
\newcommand{\pcsa}{PCSA\xspace}
\newcommand{\pcsalong}{precharge sense amplifier\xspace}

\newcommand{\averageNormalizedLatencyImprovementTacitmapOverBaselineSixtyfour}{$\sim$78$\times$\xspace}
\newcommand{\averageNormalizedLatencyImprovementOPCMOverBaselineSixtyfour}{$\sim$1205$\times$\xspace}
\newcommand{\averageNormalizedLatencyImprovementOPCMOverEPCMSixtyfour}{$\sim$15$\times$\xspace}

\newcommand{\minNormalizedLatencyImprovementOPCMOverBaselineSixtyfour}{$\sim$22$\times$\xspace}
\newcommand{\maxNormalizedLatencyImprovementOPCMOverBaselineSixtyfour}{$\sim$3113$\times$\xspace}

\newcommand{\maxNormalizedLatencyImprovementEPCMOverBaselineSixtyfour}{$\sim$154$\times$\xspace}

\newcommand{\NormalizedLatencyImprovementCIMBaselineOverGPUBaselineSixtyfourCNNFirst}{$\sim$4$\times$\xspace}
\newcommand{\NormalizedLatencyImprovementGPUBaselineOverCIMBaselineSixtyfourMLPL}{$\sim$27$\times$\xspace}

\newcommand{\averageAllNormalizedEnergyIncreaseEPCMOverBaselineSixtyfour}{$\sim$5.35$\times$\xspace}

\newcommand{\averageAllNormalizedEnergyIncreaseoPCMOverBaselineSixtyfour}{$\sim$1.56$\times$\xspace}
\newcommand{\averageAllNormalizedEnergyBoundoPCMOverBaselineSixtyfour}{60\%\xspace}

\newcommand{\averageAllNormalizedEnergyIncreaseoPCMOverEPCMSixtyfour}{$\sim$11.94$\times$\xspace}

\newcommand{\affilTUD}[0]{\textsuperscript{\S}}
\newcommand{\affilINL}[0]{\textsuperscript{$\dagger$}}

\def\BibTeX{{\rm B\kern-.05em{\sc i\kern-.025em b}\kern-.08em
    T\kern-.1667em\lower.7ex\hbox{E}\kern-.125emX}}
\begin{document}

\title{High-Performance Data Mapping for BNNs on PCM-based Integrated Photonics
}

\author{\IEEEauthorblockN{Taha Shahroodi\affilTUD, Raphael Cardoso\affilINL, Stephan Wong\affilTUD, Alberto Bosio\affilINL, Ian O'Connor\affilINL, Said Hamdioui\affilTUD }
\IEEEauthorblockA{\affilTUD dept. Quantum and Computer Engineering, Delft University of Technology, Delft, The Netherlands\\
} 
\IEEEauthorblockA{\affilINL Institut des Nanotechnologies de Lyon, École Centrale de Lyon, Lyon, France\\
}
}

\maketitle

\begin{abstract}

\sotalong (\sota) hardware implementations of \dnnlong{}s (\dnn{}s) incur high latencies and costs. \bnnlong{}s (\bnn{}s) are potential alternative solutions to realize faster implementations without losing accuracy. In this paper, we first present a new data mapping, called \effmap, suited for \bnn{}s implemented based on a \cimlong (\cim) architecture. \effmap maximizes the use of available parallelism, while \cim architecture eliminates the data movement overhead. 
We then propose a hardware accelerator based on \opcmlongsmall (\opcm) called \bnnopcmacc. \bnnopcmacc incorporates \effmap and adds an extra dimension for parallelism through \wdmlong, leading to extra latency reduction. The simulation results show that, compared to the \sota \cim baseline, \effmap and \bnnopcmacc significantly improve execution time by up to \maxNormalizedLatencyImprovementEPCMOverBaselineSixtyfour and \maxNormalizedLatencyImprovementOPCMOverBaselineSixtyfour, respectively, while also maintaining the energy consumption within \averageAllNormalizedEnergyBoundoPCMOverBaselineSixtyfour of that in the \cim baseline.

\end{abstract}

\section{Introduction} \label{sec:introduction}

Developments in \dnnlong (\dnn) in the past decade have led to significant improvements in accuracy and execution time of computer vision tasks~\cite{krizhevsky2017imagenet-ObjectRecognition1-XNORNETRef1, girshick2014rich-ObjectDetection1-XNORNETRef5, girshick2015fast-ObjectDetection2-XNORNETRef6}. However, current \dnn{} hardware implementations are relatively slow and costly to run~\cite{yuan2021comprehensive-BNNReviewBNNinEmbedded-MahdiBCIM6, courbariaux2015binaryconnect-binaryconnect-MahdiBCIM27} due to data movement overhead and expensive \gpu{}s~\cite{qin2020binary-BNNSurveyBNNinEmbedded-MahdiBCIM5}. Hence, developing high-throughput, cost-effective hardware for \dnn{}s while maintaining accuracy is critical.

Recently, researchers have proposed the use of simpler (operation-wise) and smaller \nnlong{}s (\nn{}s) such as \bnnlong{}s (\bnn{}s). \bnn{}s offer near \sotalong (\sota) accuracy on vision tasks~\cite{rastegari2016xnor-xnor_net-MahdiBCIM7} and enjoy lower memory requirements (binary values or vectors of \zeroone or \minusone)~\cite{courbariaux2015binaryconnect-binaryconnect-MahdiBCIM27, hubara2016binarized} and simplified arithmetic operations (\gatexnorop instead of multiplication or convolution)~\cite{rastegari2016xnor-xnor_net-MahdiBCIM7}. However, \bnn{}s on traditional systems using \gpu{}s incur high data movement overhead~\cite{rastegari2016xnor-xnor_net-MahdiBCIM7, chen2020phonebit-BNNGPU, nurvitadhi2016accelerating-BNNGPUCPUASICFPGA}.

The \cimlong (\cim) paradigm, especially as implemented with photonics and optical hardware, allows the data movement overhead to be alleviated while also achieving high throughput for \bnn{}s and \dnn{}s~\cite{peserico2023integrated, tait2022quantifying-PowerHungryPhaseShifters, feldmann2021parallel-photonicTensorCoreOpticalPCMIBMOPCM}. However, the previous \cim works that try to alleviate this overhead~\cite{qin2020design-BNN_ref_column1-MahdiBCIM21, hirtzlin2020digital-BNN_differential_SA-MahdiBCIM24} fail to  (fully) exploit the inherent features of the underlying hardware as they (a) lack efficient data mapping, (b) perform at most one single vector operation (e.g., \vmmlong (\vmm) or logical vector operation that is the most common operation in \nn{}s) at a time, which limits the throughput, and (c) face many design challenges such as crosstalks and large capacitances of the wiring within the memory IP of \cim, which make the design of such devices complex and limits their scalability. Even in the case of photonic \cim{}, it has been shown that at high frequencies (i.e., high noise level), recovering the result in a \cim architecture comes at a high cost and reduced accuracy~\cite{cardoso2023towards}. Fortunately, using a smaller bit count leads to an increase in the robustness, offering an opportunity for \cim architectures in photonics with high frequencies~\cite{cardoso2023towards, shahroodi2023lightspeed}.

Our paper advances the \sota \cim accelerators for \bnn{}s by providing a high throughput accelerator based on an \opcm crossbar combined with an efficient mapping method tuned to maximize the parallelism. The proposed accelerator realizes an \textbf{order of magnitude improvement in latency/throughput}. The main contributions of the paper are:

\begin{itemize} [leftmargin=*]
    \item  \effmap: A highly parallel data mapping for \bnn{} operations on any \cim design capable of performing \vmm, e.g., memristor-based crossbars based on \epcmlongsmall (\epcm) or \reramlongsmall (\reram). \effmap is designed with the conventional 1T1R memory crossbar structure in mind and is therefore compatible with many of the already evolving crossbar architectures.

    \item \bnnopcmacc: An \opcm-based \cim implementation incorporating the \effmap mapping. \bnnopcmacc ensures maximum parallelism through exploring the potential provided by the features of \cim architecture and the inherent properties of \opcm (via \wdmlong (\wdm)).

\end{itemize}

We extensively evaluate \effmap and \bnnopcmacc and compare them with \sota implementations for various \bnn{}s. Our results show that \effmap improves the latency by up to \maxNormalizedLatencyImprovementEPCMOverBaselineSixtyfour, compared to the \sota data mapping on \cim architecture for \bnn{}s. When exploiting \opcm and \effmap, \bnnopcmacc improves the latency by up to  \maxNormalizedLatencyImprovementOPCMOverBaselineSixtyfour, compared to the same baseline.

\section{Background, Related Work, and Motivation} \label{sec:background_and_relatedWork_and_motivation}

This section briefly touches on the necessary background for our work. We refer the reader to some previous works~\cite{hamdioui2017memristor, rastegari2016xnor-xnor_net-MahdiBCIM7, qin2020binary-BNNSurveyBNNinEmbedded-MahdiBCIM5, feldmann2021parallel-photonicTensorCoreOpticalPCMIBMOPCM} for detailed information.

\subsection{\cimlong for \nn{}s} 
\label{subsec:cim_pim-background_and_motivation}

\begin{figure}[htbp]
\centering
    \includegraphics[width=1\linewidth]{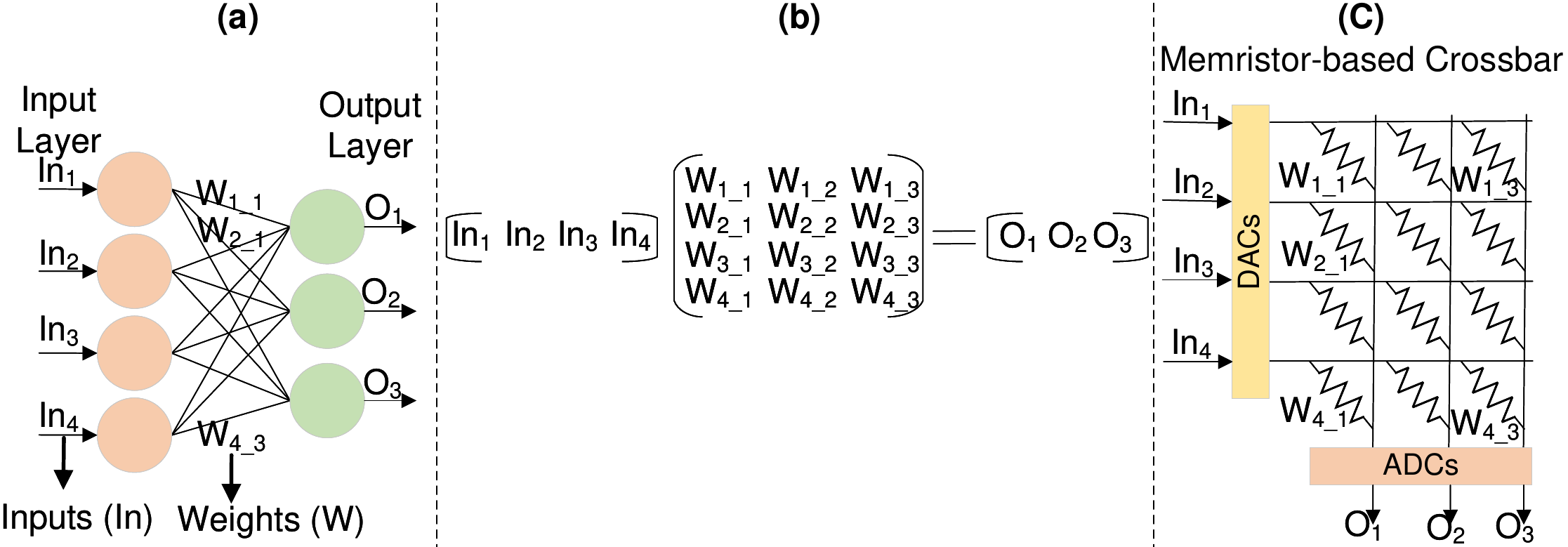}
    \caption{\cim support of \vmm{}s in \nn{}s.}
    \label{fig:cim_support_vmm_in_NN}
\end{figure}

\cimlong (\cim) is a promising computing paradigm that advocates avoiding unnecessary data movement and redesigning systems that are no longer processor-centric. Previous works~\cite{shahroodi2023RattlesnakeJake-tahamichaelshahroodiRattlesnakeJakeSAMOS2023, shahroodi2023ACaseforGenomeAnalysisWhereGenomesResides-SAMOS2023, shafiee2016isaac, ankit2019puma, shahroodi2022demeter,  shahroodi2023sievemem, ferreira2021pluto} show the potential of \cim architectures based on nanoscale emerging memory technologies for various applications ranging from \nn-based applications to those in the genomics realm. \fig{\ref{fig:cim_support_vmm_in_NN}} shows how a memristor-based crossbar supports \vmmlong (\vmm) operations, the most dominant operation in \nn{}s~\cite{shukla2018scalable-75ofNNisVMM, shahroodi2022krakenonmem, shahroodi2023swordfishMICRO}.

For a memristor-based crossbar (\fig{\ref{fig:cim_support_vmm_in_NN}}-(c)) to support \vmm, one first maps the weight matrix in \fig{\ref{fig:cim_support_vmm_in_NN}}-(b) to conductances of the memristor devices in the crossbar. Then, they apply the input vector of indexed $In_s$ as voltages to the \daclong (\dac) connected to the wordlines of each row in the crossbar. Based on Kirchhoff’s and Ohm's law, a current equivalent to accumulated current for element-wise multiplication of individual and corresponding inputs and weights in a column reaches each \adclong (\adc{}). Thus, each column performs a \maclong (\mac) operation in the analog domain, providing us with a \vmm operation across multiple columns. Since the columns can work simultaneously, the \vmm has an O(1) time complexity in this design.

\subsection{\bnnlong} 
\label{subsec:bnns_binary_neural_networks-background_and_motivation}
A \bnn{} works with binarized weights and activations (e.g., \{-1, 1\} or \{0, 1\}) instead of datatypes with higher precision, offering two advantages~\cite{courbariaux2015binaryconnect-binaryconnect-MahdiBCIM27, rastegari2016xnor-xnor_net-MahdiBCIM7}: (1) reduced storage requirement of the \nn, and (2) converting the \mac operation from high-resolution multiplication and addition to a low-cost and simpler \gatexnorop followed by an \popcount operation~\cite{rastegari2016xnor-xnor_net-MahdiBCIM7} as shown in \eque{\ref{eq:xnorpopcount-proposal_and_architecture}}. In \eque{\ref{eq:xnorpopcount-proposal_and_architecture}} $\circledast$ is convolution, $\odot$ is \gatexnorop, \popcount (or population count) of a vector or specific value is the process of finding the number of set bits (1s) in that vector/value, and vector length is the length of equally sized input ($In$) or weight ($W$) vectors.

\begin{equation}
    \label{eq:xnorpopcount-proposal_and_architecture}
    In \circledast W = 2 \times Popcount(In^{'} \odot W^{'}) - Vector\;Length
\end{equation}

Unfortunately, naively reducing both activations and weights to binary representations hurt the overall accuracy compared to high-precision (floating point or fixed point) networks. Therefore, to combat this accuracy loss, previous works~\cite{rastegari2016xnor-xnor_net-MahdiBCIM7, courbariaux2015binaryconnect-binaryconnect-MahdiBCIM27} generally follow two software-based techniques. First, tracking the updates of parameters during training via higher resolutions (floating or fixed point) while keeping the actual weights binarized. Second, using binarized activations and weights only for hidden layers and keeping the input and output layers in higher resolutions. Our proposals (\sect{\ref{sec:TacitMap_for_BNN-proposal_and_architecture}} and \sect{\ref{sec:EinsteinBarrier_architecture-proposal_and_architecture}}) also use both methods to achieve high accuracy in inference.

\subsection{\pcm-based Integrated Photonics} 
\label{subsec:PCM_based_Integrated_Photonics_OPCM-background_and_motivation}

\pcmmatlongcapital{}s (\pcm{}s) are currently the leading alternatives for non-volatile computation in silicon photonics-based platforms~\cite{peserico2023integrated}. A design that combines integrated photonics with \pcm is the commonly known \opcmlongsmall (\opcm). Compared to diffractive computing in free-space optics and previous photonic-based platforms~\cite{lin2018all-IBMOPCM29, tait2022quantifying-PowerHungryPhaseShifters}, \opcm-based designs offer \cmos-compatible manufacturing, higher speed, and lower energy consumption for the electronics interface. This is because conventional photonic-based platforms require large and power-hungry phase shifters for calibration and reconfiguration. Therefore, a design based on \opcm can reduce both the cost and the overall footprint of photonic cores for similar logical operations~\cite{cardoso2023towards}. Cardoso et al.~\cite{cardoso2023towards} showed that, with a realistic noise level, using \pcm devices in a multi-level fashion hurts the accuracy of an \opcm-based design when performing scalar multiplication. However, one can avoid this problem by using fewer levels or states in \pcm, such as using them in a binary state. In other words, the binary usage of \pcm provides the easiest solution for differentiating between the states. This fits the requirement of vectors in \bnn{}s.

One can also utilize \opcm in a \cim design, which offers three benefits compared to the same design with electronic-based \pcm as the underlying technology:

\begin{itemize}[leftmargin=*]
    \item Higher parallelization: through processing multiple vectors simultaneously using frequency space, a technique known as \wdmlong (\wdm)~\cite{feldmann2021parallel-photonicTensorCoreOpticalPCMIBMOPCM}.

    \item Higher scalability: through avoiding Joule heating, electromagnetic crosstalk, and capacitance that custom silicon computing platforms using electronic-based \pcm require~\cite{miller2017attojoule-IBMOPCM13, agrawal2017many-IBMOPCM14}.

    \item Lower design overheads and considerations: through bypassing variability, resistance drift, and cyclability challenges that affect electronic-based \cim designs~\cite{yang2009atomic-cyclability-IBMOPCM23, koelmans2015projected-PCMDrift-IBMOPCM24}.

\end{itemize}

\section{\effmap for \bnn}
\label{sec:TacitMap_for_BNN-proposal_and_architecture}

To support necessary operations in \eque{\ref{eq:xnorpopcount-proposal_and_architecture}} (e.g., \gatexnorop and \popcount) for hidden layers, we propose a data mapping for \bnn{}s, called \effmap. \effmap requires an underlying technology inherently capable of \vmm operation (\sect{\ref{subsec:cim_pim-background_and_motivation}}).

\begin{figure}[htbp]
\centering
    \includegraphics[width=1\linewidth]{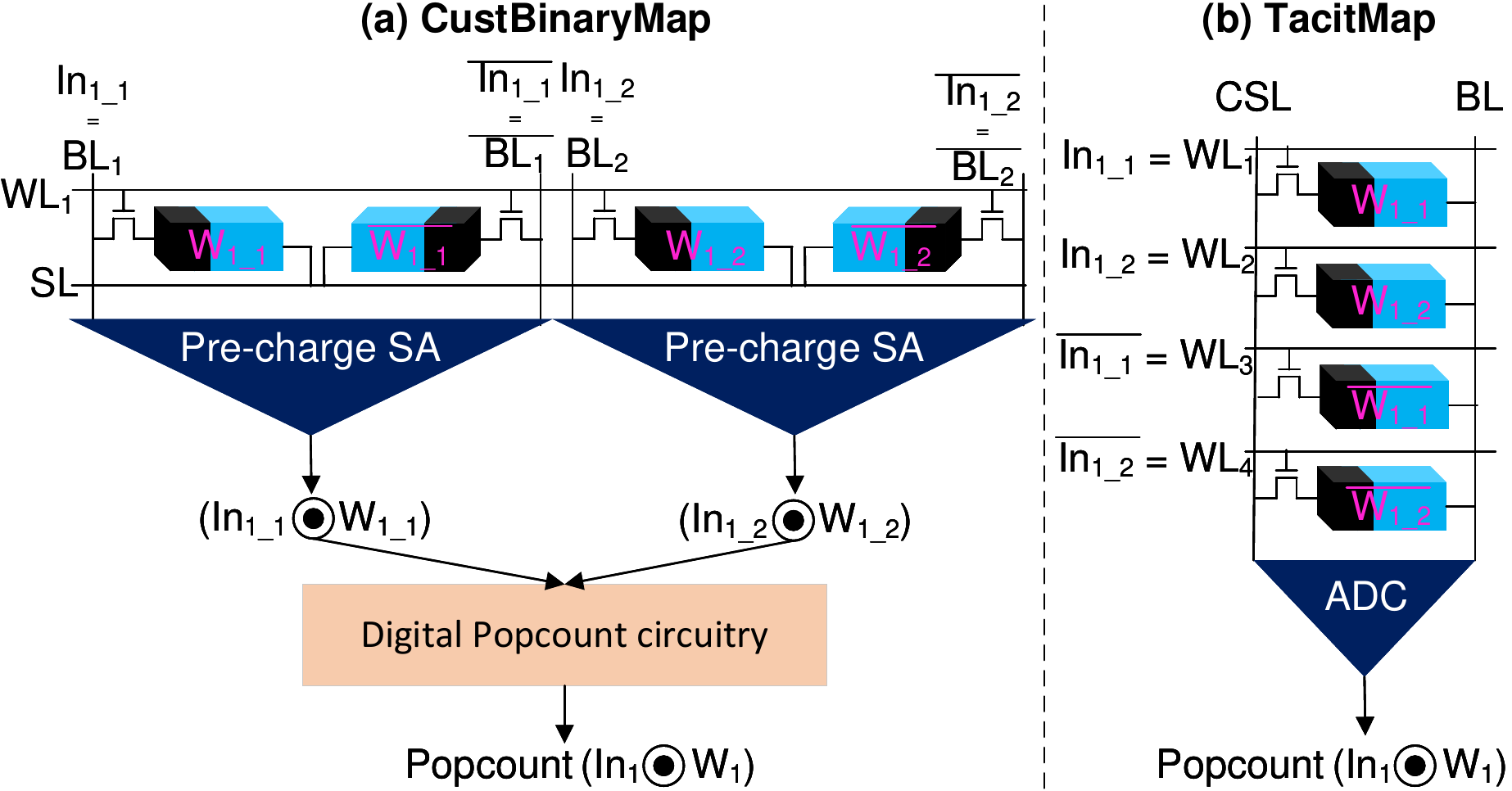}
    \caption{Concepts of \effmap vs \pcsabinarymapping~\cite{hirtzlin2020digital-BNN_differential_SA-MahdiBCIM24}.}
    \label{fig:tacitMap_concept_in2w2-TacitMap_for_BNN-proposal_and_architecture}
\end{figure}

\fig{\ref{fig:tacitMap_concept_in2w2-TacitMap_for_BNN-proposal_and_architecture}}-(a) and -(b) present a comparison between how \sota mapping (hereafter called \pcsabinarymapping)~\cite{hirtzlin2020digital-BNN_differential_SA-MahdiBCIM24} and \effmap handle a single \gatexnoroppluspopcount of \eque{\ref{eq:xnorpopcount-proposal_and_architecture}}, respectively. For a detailed description of CSL, BL, WL, and SL, please refer to previous works~\cite{chou2018n40-CSLBLWLandSL}. We assume input ($In$) and weight ($W$) vectors of length 2 bits. $In_{X\_Y}$ and $W_{X\_Y}$ represent the $Y$th bit of $X$th input and weight vector, respectively. The bar on a parameter indicates its complement value. Note that the multiplication by 2 and the subtraction in \eque{\ref{eq:xnorpopcount-proposal_and_architecture}} are constant and are implemented with minimum cost on the result of either mapping.

\pcsabinarymapping (\fig{\ref{fig:tacitMap_concept_in2w2-TacitMap_for_BNN-proposal_and_architecture}}-(a)) uses a 2T2R memory structure and places weight vectors \emph{horizontally} in memory rows. Instead of storing the weight vectors as they are, this mapping requires the programmer to interleave the weight vectors and their complements in a bitwise manner and then store the two bits $x$ and $\overline{x}$ in the two devices in the 2T2R memory cell. In contrast, \effmap (\fig{\ref{fig:tacitMap_concept_in2w2-TacitMap_for_BNN-proposal_and_architecture}}-(b)) uses a 1T1R memory structure and stores each weight vector \emph{vertically} in a column. In \effmap, instead of interleaving the weight vector with its complement, one first stores the weight vector and then, right below it, stores the complemented weight vector. Regarding the inputs, \pcsabinarymapping does the same interleaving of the input vector and its complement with the input vectors. The outputs are read through a modified \sa called \pcsalong (\pcsa), which is the \gatexnorop of the input vector and stored weight vector. Conversely, \effmap concatenates the input vector and its complement and applies it to the crossbar rows. The \gatexnoroppluspopcount is directly read out from the \adc. Although the total number of devices (i.e., memristors and transistors) is the same for both mappings, \effmap offers three main benefits compared to \pcsabinarymapping:

\begin{itemize}[leftmargin=*]
    \item 1-step \gatexnoroppluspopcount in \effmap compared to 2-step operation in \pcsabinarymapping, with no need for additional digital circuitry.

    \item Column-wise \gatexnoroppluspopcount in \effmap compared to row-wise operation in \pcsabinarymapping, enabling high parallelism for \effmap.

    \item Conventional $\mu$Arch with multiple real-world chips (i.e., 1T1R cells + \adc~\cite{ambrogio2016unsupervised-1t1rPCM1, liu202033-1t1rReRAMwithADC1}) in \effmap compared to heavily customized $\mu$Arch (i.e., 2T2R cell structure with customized \sa) in \pcsabinarymapping, making \effmap more suitable for future hardware that might be used for \bnn{}s.

\end{itemize}

\begin{figure}[htbp]
\centering
    \includegraphics[width=1\linewidth]{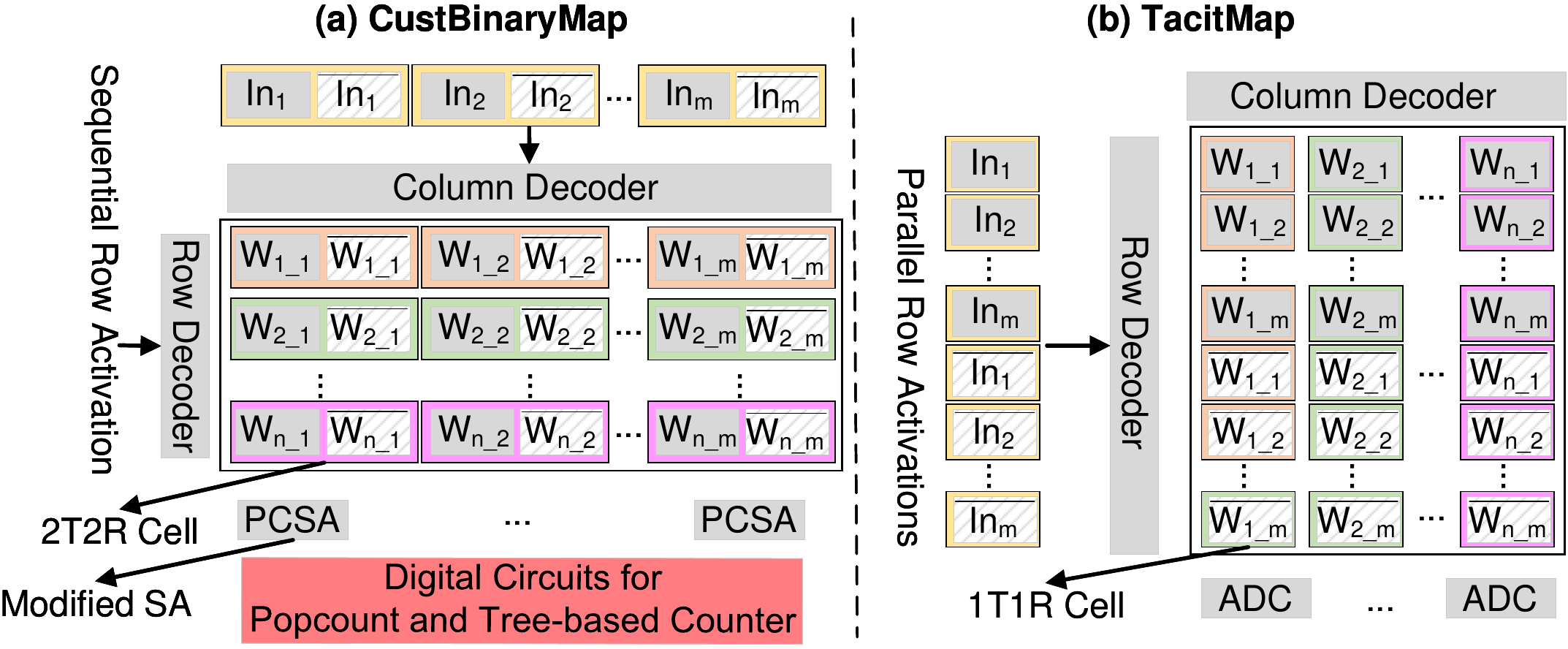}
    \caption{\effmap vs \pcsabinarymapping data mapping.}
    \label{fig:kernel_weight_mapping_BinaryTraditionalVsTacitMap_Horizontal-proposal_and_architecture}
\end{figure}

\fig{\ref{fig:kernel_weight_mapping_BinaryTraditionalVsTacitMap_Horizontal-proposal_and_architecture}} presents \effmap against \pcsabinarymapping at the crossbar level. We observe that \effmap enables the crossbar to perform  $n$ \gatexnoroppluspopcount via a single \vmm operation in only 1 step and reads the results from \adc{}s simultaneously. In contrast, \pcsabinarymapping takes a minimum of $n$ steps for the same operations because  \pcsabinarymapping utilizes \pcsa to perform the logical \gatexnorop for one input and one weight vector of size $m$ and processes $n$ weight vectors sequentially in $n$ steps. Moreover, \pcsabinarymapping also needs to perform post-processing on the read output on every final vector using two additional digital components: (1) a fully digital five-bit counter per crossbar column for local \popcount and (2) a tree-based \popcount circuit per several connected crossbars for a global \popcount. Theoretically, using the same underlying device, \effmap should achieve up to $n\times$ lower execution time compared to \pcsabinarymapping.

\begin{figure*}[htbp]
\centering
    \includegraphics[width=1\linewidth]{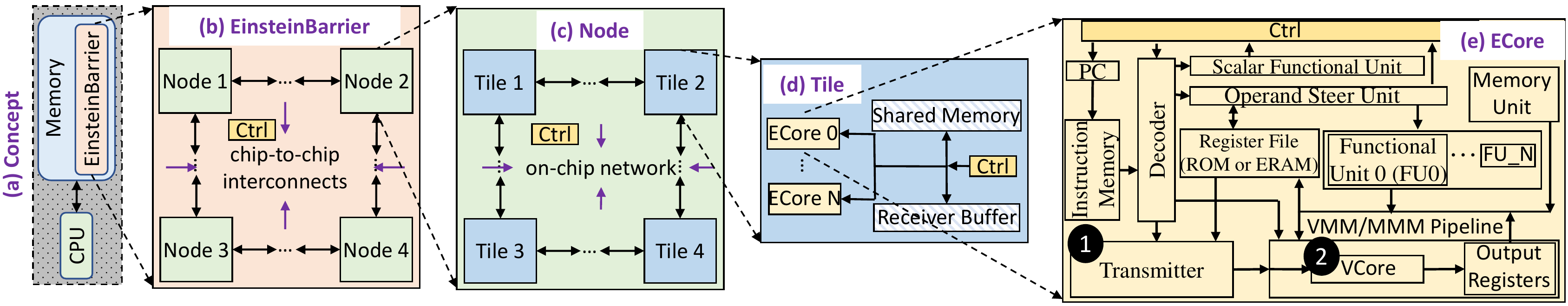}
    \caption{\bnnopcmacc system placement and overview.}
    \label{fig:Einstein_Barrier_System_Overview_Horizontal-proposal_and_architecture}
\end{figure*}

Note that \effmap is compatible with any technology for the crossbar that supports \vmm, e.g., \epcm- or \opcm-based crossbars.

\section{\bnnopcmacc Architecture}
\label{sec:EinsteinBarrier_architecture-proposal_and_architecture}

\fig{\ref{fig:Einstein_Barrier_System_Overview_Horizontal-proposal_and_architecture}}-(a) presents an overview of the \bnnopcmacc concept and its system placement. We envision \bnnopcmacc as an accelerator that is part of the memory itself. \bnnopcmacc is a spatial architecture (similar to \puma~\cite{ankit2019puma}) with four levels: Nodes, Tiles, External Cores (\ecore{}s), and \vmm-enabled cores (\vcore{}s). \bnnopcmacc extends the ISA discussed in a earlier work~\cite{ankit2019puma} to support multiple simultaneous \vmm{}s, called \mmmlong (\mmm) hereafter. 

\bnnopcmacc hierarchical organization, Tile architecture, and \ecore{}s provide the generality and reconfigurability needed to support various \bnn{}s and various technologies. The \ecore{} and the new $\mu$Arch required for \gatexnoroppluspopcount brings the generality needed to support \effmap and multiple technologies in \vcore{}s as long as they support \vmm operation in a crossbar. The \ecore{} and \vcore designs prepare \bnnopcmacc in particular to adopt \opcm technology and its advantages. Finally, by simply adopting \cmos-compatible \opcm-based \vcore{}s, \bnnopcmacc avoids the challenges of \epcm-based \cim architecture (\sect{\ref{subsec:PCM_based_Integrated_Photonics_OPCM-background_and_motivation}}).

\subsection{\opcm-based \wdm-enabled \ecore}
\label{subsec:opcm_core_VCore-proposal_and_architecture}

\bnnopcmacc utilizes integrated photonics with \pcm devices in the crossbar. This choice of \opcm-enabled \ecore{}s demands two specific additional components, namely \vcore{} and transmitter, compared to other \cim-enabled designs. This choice provides an extra dimension for parallelization through \wdm and avoids Joule heating and resistance drift in electronic emerging memories~\cite{miller2017attojoule-IBMOPCM13, joshi2020accurate-deviceVariability-IBMOPCM22}.

\subsubsection{\vcore structure}
The \opcm-based \vcore consists of a memory crossbar (i.e., a tile) with each cell in the crossbar being a single \pcm device capable of storing 1 bit of data (binary \pcm discussed in \sect{\ref{subsec:PCM_based_Integrated_Photonics_OPCM-background_and_motivation}}). A tile also includes all the necessary peripheries for read and write operations (e.g., DACs and \adc{}s). \bnnopcmacc adds 1 more component to the readout circuitry of the \opcm core: \tialong (\tia) on the output (receiver). \bnnopcmacc uses \tia to feed \adc{}s, acting as a deserialization stage in the output.

\subsubsection{\wdm}
\begin{figure}[htbp]
\centering
    \includegraphics[width=1\linewidth]{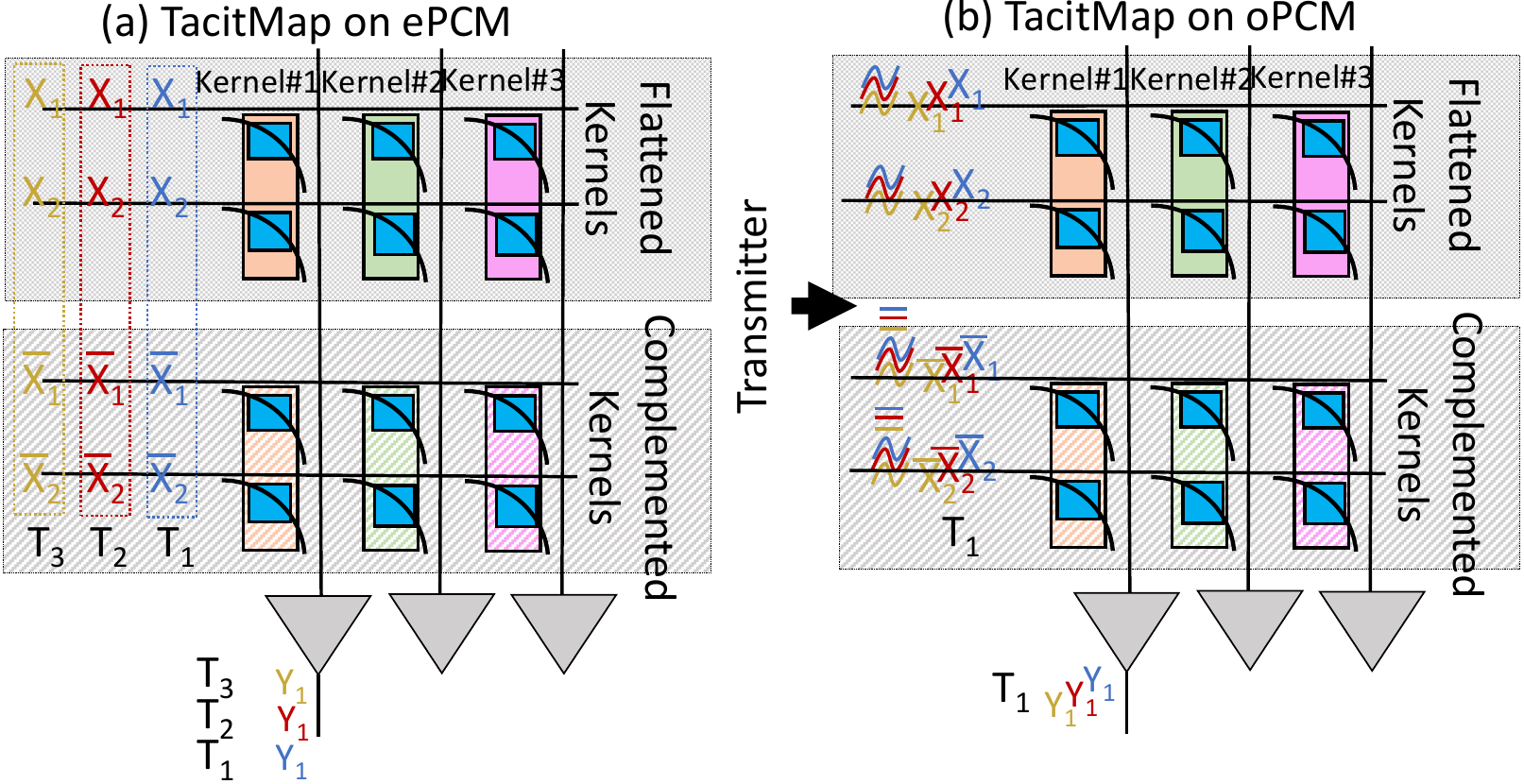}
    \caption{\wdm in \opcm core.}
    \label{fig:wdm_walk_through-proposal_and_architecture}
\end{figure}

\fig{\ref{fig:wdm_walk_through-proposal_and_architecture}} depicts an example to present the concept and benefits of \wdm in an \opcm core. We assume 3 2-bit activation vectors of $X\_{i}$ distinguished by vectors with yellow, red, and blue colors. The indices refer to the bit number. Moreover, we assume 3 2-bit kernel vectors of $k\_{i,j}$, where $i$ denotes the activation/kernel vector and $j$ denotes the bit position in that activation/kernel vector. Each of these kernel vectors is grouped in a box of orange, green, or pink color. The complements use the same color but striped boxes.

In \fig{\ref{fig:wdm_walk_through-proposal_and_architecture}}, \effmap requires three columns and 4 (2$\times$2) rows of the crossbar to store the kernels and their complement. \effmap requires three \vmm operations\footnote{For simplicity, we assumed that the columns could be read out in parallel and they do not share an \adc{}. We will revisit this in \sect{\ref{sec:evaluation_methodology-evaluations_and_results}}.} to process all the required \gatexnoroppluspopcount{}s; i.e., 1 \vmm per each activation vector, wherein each \vmm we use the vector of that activation concatenated by its complement as the input to the crossbar.

\fig{\ref{fig:wdm_walk_through-proposal_and_architecture}}-(a) depicts the scenario for a conventional \epcm-based \vmm-enabled crossbar. In this case, the required \vmm{}s happen in consecutive time-steps, denoted by $T_1$, $T_2$, and $T_3$ in \fig{\ref{fig:wdm_walk_through-proposal_and_architecture}}-(a). The input vector size in this scenario is 4, the number of inputs is 3, and the matrix is $4\times3$.

On the other hand, \fig{\ref{fig:wdm_walk_through-proposal_and_architecture}}-(b) depicts the same scenario but for an \opcm-based \vcore. Using an optical transmitter that we discuss next, one can combine our 3 input vectors together into a single input and feed that single input to the crossbar. Therefore, only 1 time-step, i.e., $T_1$, is required to finish the operation. Here, the input vector and the matrix size are still 4 and $4\times3$, respectively. However, the number of input vectors is reduced to 1. Therefore, effectively, \wdm-enabled an \mmm of size $4\times4\times3$. We call the number of wavelengths that can be combined into a single wavelength and still be detectable later (with acceptable noise in \tia) the \wdm capacity. Current technologies can support up to a capacity of $K=16$~\cite{feldmann2021parallel-photonicTensorCoreOpticalPCMIBMOPCM}, meaning a theoretical 16$\times$ improvement in performance compared to \epcm.

\subsubsection{Transmitter structure}
\begin{figure}[htbp]
\centering
    \includegraphics[width=0.90\linewidth]{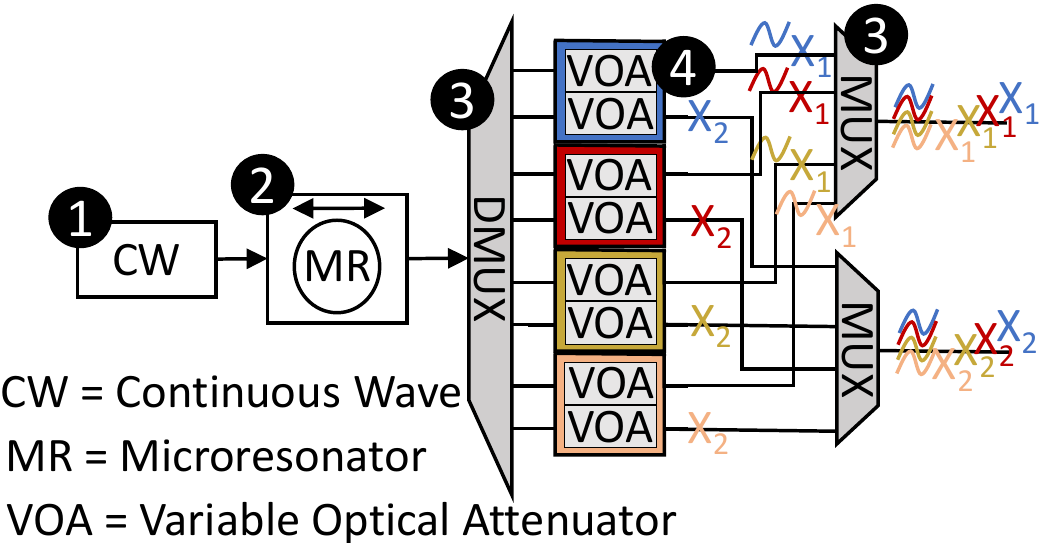}
    \caption{Transmitter overview.}
    \label{fig:transmitter_Einstein_Barrier-proposal_and_architecture}
\end{figure}

To support optical inputs and \wdm, \bnnopcmacc adds a transmitter circuit (\circled{1} in \fig{\ref{fig:Einstein_Barrier_System_Overview_Horizontal-proposal_and_architecture}}) at the \ecore{} level feeding the \vmm/\mmm pipeline, where the actual \opcm-based core (\circled{2}) resides. \fig{\ref{fig:transmitter_Einstein_Barrier-proposal_and_architecture}} presents a high-level overview of the transmitter circuit and components.

The transmitter has four main components: \circled{1} a laser to provide a single-wavelength continuous wave beam, \circled{2} a microresonator-based optical frequency comb to concentrate the optical power and excite new wavelengths based on non-linearities, \circled{3} DMUXs/MUXs for feeding individual waves to each \voalong (\voa) and creating a single wave carrying information on multiple bits from different vectors, and \circled{4} \voa{}s to encode the information of each input into waves via changing the amplitude. In \fig{\ref{fig:transmitter_Einstein_Barrier-proposal_and_architecture}}, the transmitter combines four 2-bit vectors (of different colors) into a single vector of 2-bit width.

\subsection{\opcm-based \ecore Overheads}
\label{subsec:opcm_core_overhead-proposal_and_architecture}

We showed (\sect{\ref{subsec:opcm_core_VCore-proposal_and_architecture}}) that using \opcm provides higher parallelism (simultaneous \vmm{}s vs. single \vmm) for the same vector operations via \wdm. However, this extra parallelism comes at the cost of power for the additional components. Assume a core with \wdm capacity of K and crossbars of size $M\times N$. Such a crossbar incurs an extra power modeled by \eque{\ref{eq:power_overhead_crossbars_wdm_opcm-proposal_and_architecture}}, where N is \# of \tia{}s, each of which consumes \SI{2}{\milli\watt}.

\vspace{-0.3cm}
\begin{equation}
    \label{eq:power_overhead_crossbars_wdm_opcm-proposal_and_architecture}
    P_{crossbar} = N \times \SI{2}{\milli\watt}
\end{equation}

The transmitter power overhead is presented in \eque{\ref{eq:power_overhead_transmitter_wdm_opcm-proposal_and_architecture}}, where it accounts for the required power for the laser, modulators, and tuning~\cite{cardoso2022energy-powerRaphaelCalc2}.

\vspace{-0.5cm}
\begin{equation}
    \label{eq:power_overhead_transmitter_wdm_opcm-proposal_and_architecture}
    P_{total} = P_{laser} + 3 \times KM \SI{}{\milli\watt}+ \frac{3 \times KM + 1}{k} \times \SI{45}{\milli\watt}.
\end{equation}

\section{Evaluation Methodology}
\label{sec:evaluation_methodology-evaluations_and_results}

\subsection{Implementations and Models}
We implement \bnnopcmacc as a heavily extended version of \puma architecture and compiler~\cite{pumasimulator2019}. This implementation\footnote{We open-source our experimental setup upon acceptance.} accounts for (1) \wdm capability of \opcm cores, (2) new configurations related to integrated photonics, and (3) power and area overheads introduced by extra components of \opcm cores, e.g., laser. For the photonics components, we use our device-aware extended circuits~\cite{zrounba2021exploration, cardoso2022energy-powerRaphaelCalc2, feldmann2021parallel-photonicTensorCoreOpticalPCMIBMOPCM, polster2016efficiency-powerRaphaelCalc1}. Our \epcm-based crossbars are based on extensive characterization done in the EU project MNEMOSENE project and previous works~\cite{MNEMOSENE-TUDelftEU, feldmann2021parallel-photonicTensorCoreOpticalPCMIBMOPCM}, generously provided to us by the partners. To evaluate additional CMOS circuitry of our design (e.g., such as MUXs), we use Synopsys Design Compiler and synthesize them in the target technology to obtain their execution time, power, and area. We apply the prominent technology scaling rules~\cite{sarangi2021deepscaletool} to the configuration numbers of \puma architecture to ensure our design components are based on the same technology node. 

\subsection{Designs and Baselines}
We evaluate the effectiveness of \effmap and \bnnopcmacc separately using two different configurations: (1) \epcmeffmap that is \effmap on electronic \pcm-based cores, and (2) \bnnopcmacc that still uses \effmap but utilizes \opcm-based \vcore{}s. We use two baselines: (1) the design in \cite{hirtzlin2020digital-BNN_differential_SA-MahdiBCIM24}, a \sota hardware accelerator for \bnn{}s, hereafter called \baseline, and (2) a \gpu implementation of the same network (called \gpubaseline). We use the same \pcm configuration in \epcmeffmap for \baseline.

\subsection{Networks and Datasets}
We evaluate all designs over 6 \bnn{}s (3  convolutional networks and 3 multilayer perceptrons (MLPs)) with various sizes from MlBench~\cite{chi2016prime}. We use MNIST and CIFAR-10 for the datasets. Note that neither \effmap nor \bnnopcmacc affect the accuracy of target \bnn{}s and simply accelerates them via efficiently handling their \gatexnoroppluspopcount in parallel. 

\section{Evaluation Results and Discussions} \label{sec:evaluations_and_results}

\subsection{Performance Analysis}
\label{subsec:performance-evaluation_results-evaluations_and_results}

\fig{\ref{fig:latency_Improvements_all_networks_BaselineePCM_ePCMMapping_OPCMMapping_64-evaluations_and_results}} presents the latency improvement of \epcmeffmap and \bnnopcmacc normalized to \sota. The y-axis uses a log scale. We make four key observations:

\begin{figure}[htbp]
\centering
    \includegraphics[width=1\linewidth]{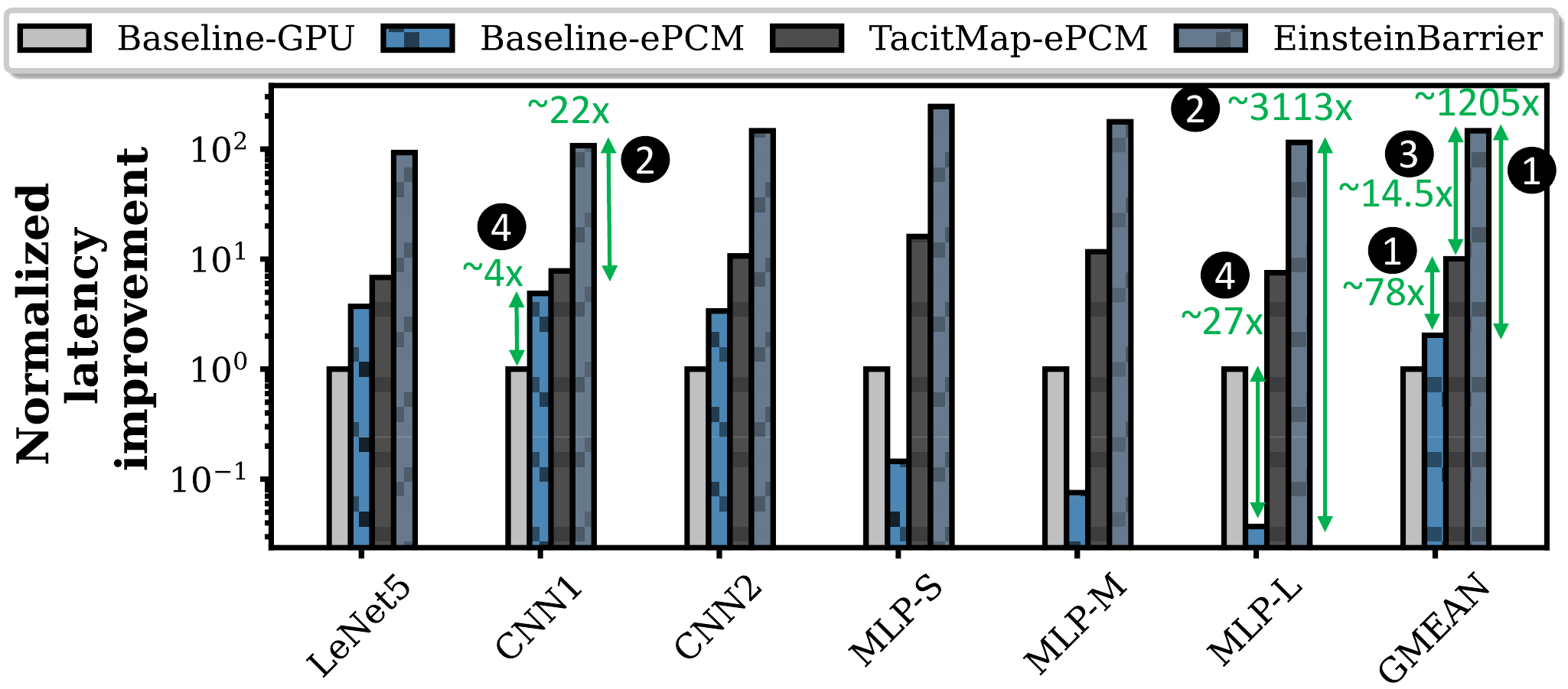}
    \caption{Normalized latency improvements over all networks.}
    \label{fig:latency_Improvements_all_networks_BaselineePCM_ePCMMapping_OPCMMapping_64-evaluations_and_results}
\end{figure}

\begin{itemize}[leftmargin=*]
    \item Both \epcmeffmap and \bnnopcmacc improve the latency over \baseline irrespective of the underlying network. On average, \epcmeffmap and \bnnopcmacc improve the performance by \averageNormalizedLatencyImprovementTacitmapOverBaselineSixtyfour and \averageNormalizedLatencyImprovementOPCMOverBaselineSixtyfour, respectively (\circled{1}). These are because, unlike the \baseline, \epcmeffmap and \bnnopcmacc parallelize \gatexnorop with \popcount and parallelize many \gatexnoroppluspopcount{}s via the proposed vertical data mapping.

    \item The latency improvement is network-dependent and varies from \bnn to \bnn. Specifically, the latency improvements over \baseline vary from \minNormalizedLatencyImprovementOPCMOverBaselineSixtyfour to \maxNormalizedLatencyImprovementOPCMOverBaselineSixtyfour for \bnnopcmacc (\circled{2}). This happens due to (1) the relation between the size of the hidden layers (binary layers) and the first and last layers and (2) available parallelism in the \gatexnoroppluspopcount operations of each \bnn. In the evaluated \bnn{}s, larger \bnn{}s contain more parallel \gatexnoroppluspopcount operations.

    \item \bnnopcmacc improves the latency on average \averageNormalizedLatencyImprovementOPCMOverEPCMSixtyfour (\circled{3}) with the exact data flow compared to \epcmeffmap. This happens due to the extra parallelism dimension enabled by \wdm and the fast crossbar read of \opcm core. This is while the improvement is still network-dependent. Unfortunately, the achieved improvement due to the technology is still lower than the \wdm capacity (i.e., $K=16$). This is simply due to the underlying network, and we expect it to increase for larger networks. We leave this exploration as future work.

    \item \baseline does not always improve the latency over a \gpubaseline. For example, see \circled{4} in \fig{\ref{fig:latency_Improvements_all_networks_BaselineePCM_ePCMMapping_OPCMMapping_64-evaluations_and_results}}, while \baseline is \NormalizedLatencyImprovementCIMBaselineOverGPUBaselineSixtyfourCNNFirst faster than \gpubaseline for our first \cnn, it is \NormalizedLatencyImprovementGPUBaselineOverCIMBaselineSixtyfourMLPL slower than \gpubaseline for our MLP-L network. This happens since in some networks, such as our MLP workloads, \baseline has to serialize \gatexnoroppluspopcount compared to \gpubaseline, to the extent that the benefits for reducing the data movement overhead diminish.

\end{itemize}

\subsection{Energy Analysis}
\label{subsec:power_or_energy-evaluation_results-evaluations_and_results}

\fig{\ref{fig:energy_Improvements_all_networks_BaselineePCM_ePCMMapping_OPCMMapping_64-evaluations_and_results}} compares the energy consumption of \epcmeffmap and \bnnopcmacc normalized to the \baseline. The y-axis is in a log scale. We make two key observations:

\begin{figure}[htbp]
\centering
    \includegraphics[width=1\linewidth]{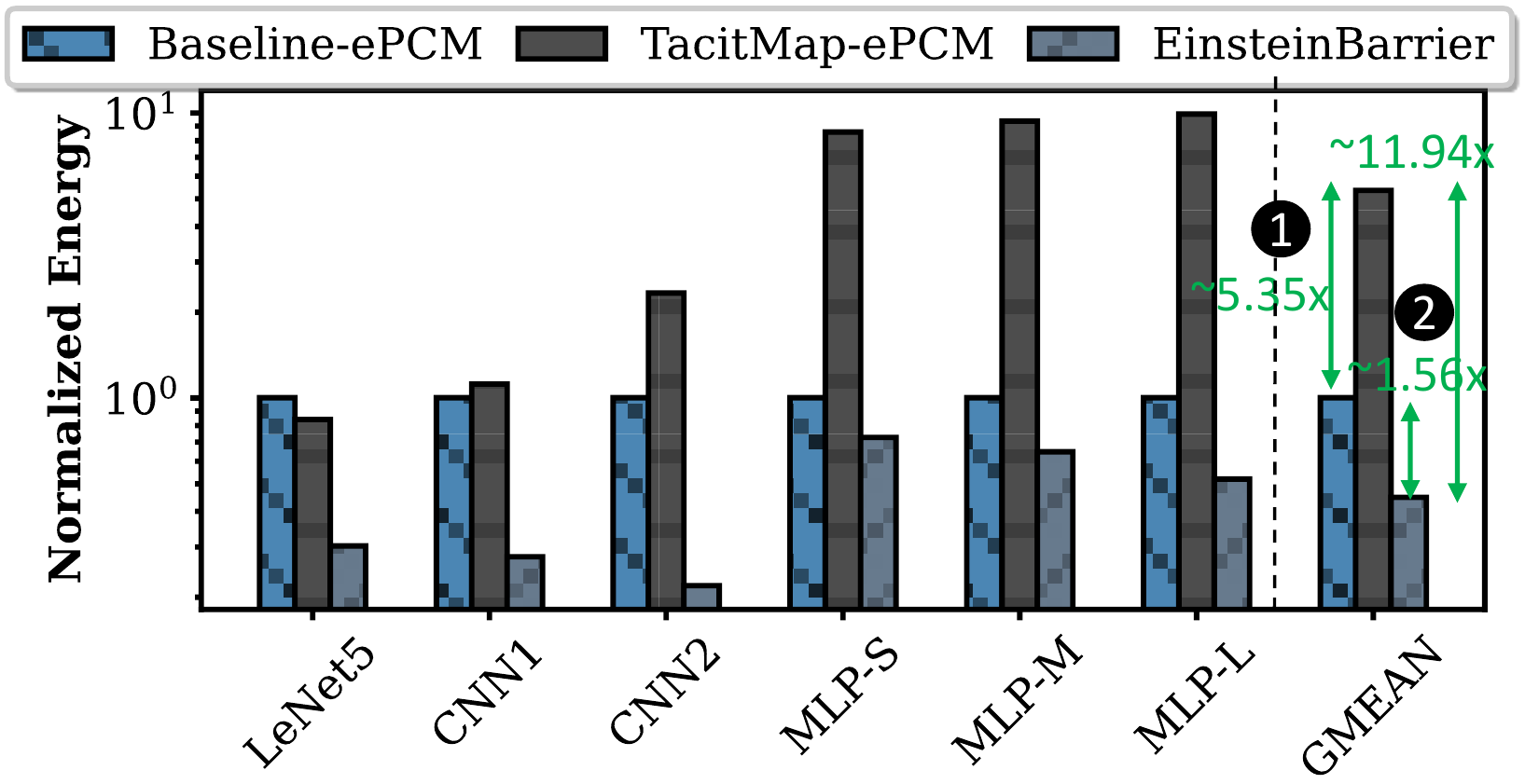}
    \caption{Normalized energy consumption over all networks. }
    \label{fig:energy_Improvements_all_networks_BaselineePCM_ePCMMapping_OPCMMapping_64-evaluations_and_results}
\end{figure}

\begin{itemize}[leftmargin=*]
    \item  On average, \epcmeffmap increases the energy consumption compared to \baseline by \averageAllNormalizedEnergyIncreaseEPCMOverBaselineSixtyfour, as \epcmeffmap requires power-hungry \adc{}s while \baseline uses \sa{}s (\circled{1}).

    \item On average, \bnnopcmacc improves the energy consumption by \averageAllNormalizedEnergyIncreaseoPCMOverBaselineSixtyfour and \averageAllNormalizedEnergyIncreaseoPCMOverEPCMSixtyfour over \baseline and \epcmeffmap, respectively (\circled{2}). This is achieved because \bnnopcmacc requires a lower number of crossbar activations by computing multiple outputs at the same time while using the same crossbar, ADCs, and other peripheries.

\end{itemize}

\subsection{Discussions and Future Works} 
\label{sec:discussions_limitations_futureWorks}

\noindent
\textbf{Multi-Level \pcm Devices}.
Our work uses \pcm{}s in a binary mode. However, recent works~\cite{feldmann2021parallel-photonicTensorCoreOpticalPCMIBMOPCM, zhang2019miniature-IBMOPCM65} show the potential for multi-bit devices at the cost of increased noise. We leave extending \effmap on multi-bit cells for future work.

\noindent
\textbf{Design Space Exploration of \opcm-based \vcore{}s}.
We evaluated \bnnopcmacc using fixed laser, array sizes, and other system configurations due to our limited access to accurate specs of different components (particularly those needed in the transmitter). A study that can freely explore this design space is encouraged and left for future work.

\section{Conclusion} \label{sec:conclusion}

This paper proposes an efficient data flow for \bnn{}s, \effmap, and a \cmos-compatible \opcm-based hardware accelerator based on integrated photonics principles, called \bnnopcmacc, to fully exploit the possible parallelism with \effmap. Our latency and energy evaluations suggest an average improvement of  \averageNormalizedLatencyImprovementOPCMOverBaselineSixtyfour and \averageAllNormalizedEnergyIncreaseoPCMOverBaselineSixtyfour, respectively, for \bnn{}s on \opcm-based accelerators. This is the first step towards an optimized and efficient hardware realization for \bnn{}s using these emerging technologies. Hence, our work encourages further investigations of \opcm in the \nn realm.

\bibliographystyle{IEEEtran}
{\footnotesize
\bibliography{main}}

\end{document}